# Exponential decrease of Ramsey linewidth via coherent atomic phase tracking

Gongxun Dong[1,2], Jinda Lin[1*], Jianliao Deng[1], and Yuzhu Wang[1,3]

[1] *Key Laboratory of Quantum Optics and Center for Cold Atom Physics，Shanghai Institute of Optics and Fine Mechanics, Chinese Academy of Sciences, Shanghai, 201800, China*

[2] *University of Chinese Academy of Sciences, Beijing, China*
[3] *yzwang@mail.shcnc.ac.cn*
*Corresponding author: Jinda@siom.ac.cn

We propose an effective method to decrease the Ramsey linewidth of microwave atomic clock by tracking the coherent atomic phase through nondestructive measurement. The free evolution time T between two Ramsey pulses is divided into *N* subsections. In each subsection, the coherent phase of the atomic ensemble is probed by an additional large-detuning laser. The probed phase in the previous subsection is used to change the phase of the microwave pulse in the following subsection. After tracking the coherent atomic phase for *N*-1 times, the final linewidth of Ramsey fringe decreases quickly as $N/(2^N-1)$. We then demonstrate this proposal in a vapor cell atomic clock. The linewidth of the Ramsey fringe is decreased down to 27 Hz after 5 successive sub-interrogations, which is much narrower than the linewidth of 142 Hz obtained by normal Ramsey spectroscopy.

*PACS number:*

The precise realization of the second was made by the development of frequency standards based on the energy transitions of atoms. In the 1930s, Isidor Rabi invented the magnetic resonance technique [1] and suggested to make a clock of extraordinary accuracy with the measurement of the natural resonant frequencies of atoms. In 1949, Norman Ramsey invented the separated oscillatory field method [2] to interrogate the atoms with two microwave pulses and made it possible to build a much more accurate clock. Since then, Ramsey method has been utilized by modern primary frequency standards. [3] The linewidth $\Delta\nu$ of the Ramsey fringe is determined by the interrogation time *T* between the two microwave pulses with $\Delta\nu = 1/2T$ [4], which directly determines the clock stability characterized by $\sigma(\tau) = \frac{1}{\pi \cdot SNR \cdot \frac{\nu_0}{\Delta\nu}}\sqrt{\frac{T_c}{\tau}}$ [4], where $T_c$ is the cycle time, *SNR* is the signal-to-noise ratio, $\frac{\nu_0}{\Delta\nu}$ is the atomic quality factor, and $\tau$ is the averaging time.

Extending the interrogation time of the atomic ensemble, i.e., decreasing the linewidth $\Delta\nu$, is an efficient way to improve the clock stability. The introduction of buffer gases increases the coherence time of the vapor cell atomic clock to about several milliseconds [5]. The use of an atomic fountain [6] with cooled atomic cloud allows for coherence times on the order of one second. With the operation of artificial rephaing through a sequence of microwave pulses [7], the spin-self-rephasing [8,9], or the atomic phase lock (APL) [10], the coherence time on the microwave clock transition was extended up to tens of seconds. Moreover, the APL are proposed to improve the stability of clocks much quickly from $1/\tau^{1/2}$ to $1/\tau$ [11, 12], which is also demonstrated by using the zero dead time operation of interleaved atomic clock [13].

In this paper, we propose an effective method to exponentially decrease the Ramsey linewidth of microwave atomic clock without using any operation to extend the coherence time. On the contrary, we divide the interrogation time *T* into *N* successive sub-interrogation with time *T/N*. After each sub-interrogation, the evolve phase $\varphi_i$ of the atomic ensemble is probed by nondestructive measurement with an additional large-detuning laser. $\varphi_i$ is used to change the phase of the second microwave pulse in the following sub-interrogation. At the final projective measurement, we obtain the linewidth of Ramsey fringe with $N/(2^N-1) \cdot 1/2 T$. We verify the proposal in a vapor cell atomic clock with mixer buffer gases and obtain a narrow linewidth of Ramsey fringe of 27 Hz.

We analyze the Ramsey method by using the Bloch representation [4,11,14] where the two level system interacting with an electromagnetic field is visualized as a spin-1/2 particle in a magnetic field. As shown in Fig.1, *u, v,* and *w* are the three components of the Bloch vector. Steps (a)-(d) depict the conventional Ramsey procedure. (a) Prepare the atoms in |↓⟩ state by resonant optical pumping. (b) A π/2 microwave pulse with 0° phase brings the atoms into a superposition state of |↓⟩ and |↑⟩. (c) Free precession of time *T* between two microwave pulses. The frequency difference $\Delta\nu$ between the clock transition and the microwave field leads to a relative phase of $\varphi = 2\pi\Delta\nu T$. (d) Apply a second π/2 microwave pulse with 0° phase. At the projective measurement procedure, $\varphi$ is mapped onto the probability *P* of finding the atom in the |↑⟩ state. The probability shows a cosine variation as the function of $(\cos\varphi + 1)/2$ near resonance. The full width at half maximum of the Ramsey fringe is given by *P* =1/2 or $\varphi = \pm\pi/2$, leading to the linewidth of $\Delta\nu = 1/2T$.

Instead of mapping the precession phase $\varphi = 2\pi\Delta\nu T$ directly onto the probability of population, we propose a new method based on atomic phase tracking to accelerate the final precession phase $\varphi_f$. The procedure is different with the conventional Ramsey method from step (c) and is structured as follows. We divide the precession time *T* in (c) into *N* successive sub-interrogation with time *T/N*.

(a) Initialize the Bloch vector along –*w* axis,
(b) A π/2 microwave pulse with 0° phase rotates the Bloch vector around -*v* axis by π/2,

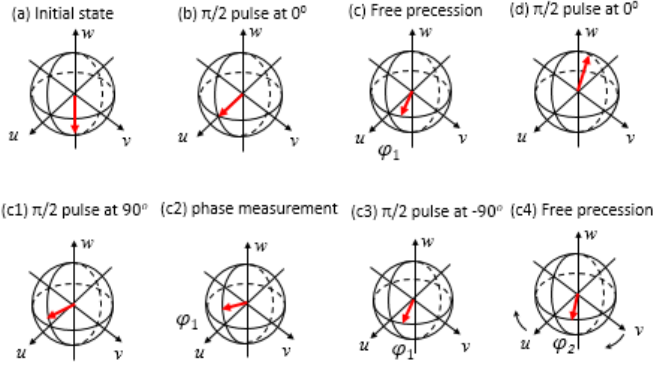

FIG.1 (color online) Bloch sphere representation of Ramsey method. (a) Initialize the Bloch vector along –$w$ axis; (b) Rotate the Bloch vector around -$v$ axis by $\pi/2$ along $u$ axis; (c) Bloch vector evolves on the equatorial plane for time T with $\varphi = 2\pi\Delta\nu T$ relative to $u$ axis. (d) Rotate the Bloch vector around –$v$ axis by $\pi/2$ and estimate the $w$ component. (c1) Without doing step (d), after evolving time for $T/N$, rotate the Bloch vector around $u$ axis by $\pi/2$ into the $u$-$w$ plane, (c2) Estimate the evolution phase by nondestructive measurement with $\varphi'_1$; (c3) Rotate backward the Bloch vector around -$u$ axis by $\pi/2$; (c4) Bloch vector continues to evolve on the equatorial plane with $\varphi_2$ and rotate the $u$-$v$ axis clockwise around $w$ axis by angle $\varphi'_1$. Repeat steps (c1)-(c4) for ($N$-1) times. In the $N$th precession, rotate the $u$-$v$ plane around $w$ axis by angle $\varphi'_{N-1}$, and estimate the $w$ component by projective measurement with $\varphi_N$.

(c) Bloch vector evolves on the equatorial plane for T/N with $\varphi_1 = 2\pi\Delta\nu T/N$ relative to $u$ axis,

(c1) A $\pi/2$ microwave pulse with 90° phase rotates the Bloch vector around $u$ axis by $\pi/2$,

(c2) Measure the evolution phase $\varphi_1$ without destroying the atomic coherence and obtain $\varphi'_1$

(c3) A $\pi/2$ microwave pulse with -90° phase rotates the Bloch vector back to step (c),

(c4) Bloch vector continues to evolve on the equatorial plane for $T/N$. Rotate $u$-$v$ axis around $w$ axis to $u(\varphi'_1)$-$v(\varphi'_1)$ by changing the phase of the following microwave pulse to 90-$\varphi'_1$.

Repeat steps (c1) to (c4) for $N$ times. During the of the $i$th precession, the nondestructive measurement should give the result of $\varphi'_i=2\varphi'_{i-1}+\varphi_i$. where $\varphi'_1 = \varphi_1$, $\varphi'_2 = 3\varphi_1$, $\varphi'_3 = 7\varphi_1$, …, $\varphi_f = (2^N - 1)\varphi_1$. At the final step, the projective measurement of the precession phase $\varphi_f$ determines the linewidth of the Ramsey fringe, where $\varphi_f = \pm\pi/2$ corresponds to $P = 1/2$. The linewidth is then determined by $(2^N - 1) \cdot 2\pi\Delta\nu T/N = \pi$,
i.e. $\Delta\nu = N/(2^N - 1) \cdot 1/2T$.

We verify the proposal in the optically pulsed pump [5,15] vapor cell atomic clock. The experimental setup and the time sequences are shown in Fig.2. A pair of polarizers (P and A) with extinction ratio of about 100000:1 are placed orthogonally at both sides of the physical package for the dispersive detection [16]. The dispersive detection scheme becomes necessary here for the reason that the dispersive detection can probe the atomic phase without destroying the atomic coherence. The physical package is similar with the experimental setup in ref. [16]. The atomic vapor cell is filled with $^{87}$Rb atoms and mix buffer gases of $N_2$ and Ar.

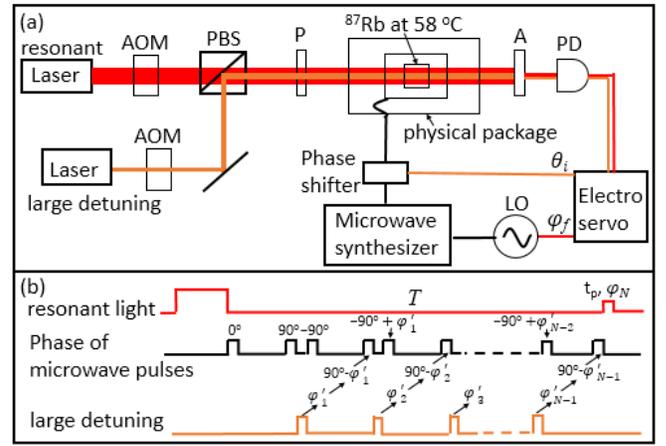

FIG.2 (color online) Experimental setup (a) and time sequences (b) for atomic phase tracking. (a) A large detuning laser (orange line) is used to probe the atomic phase during the free evolution stage in the conventional pulsed optically pumped atomic clock. The probe phase is used to change the phase of the microwave pulse as shown in (b). AOM-acoustic optical modulator, PBS-polarizer beam splitter, P-polarizer, A-analyzer, PD-photodiode detector, LO-local oscillator. (b) The resonant laser (red line) is used for optical pumping at the beginning and projective detection at the final precession. The large detuning light (orange line) is used for nondestructive measurement of the atomic precession phase $\varphi'_i$ during the free evolution. The phases of microwave pulses (black line) are changed according to $\varphi'_i$. T=3.5 ms, $t_m$=50 μs, and $t_p$=50 μs.

The atomic vapor is heated to 58 °C and the coherent time of the atomic ensemble is about 4 ms. Three layers of magnetic shielding made of μ-metal isolate the cell from the outside magnetic field and a solenoid with central magnetic field of 15 mGs provides the quantized axis. Compared to the conventional optically pulsed pump atomic clock, we need two laser in the experiment. The resonant light locked to the $|F_g=2\rangle$—$|F_e=2\rangle$ transition of $^{87}$Rb $D_1$ line through saturated absorption spectrum is used for optical pumping and projective measurement. The other one with large detuning of about 1 GHz to the resonant one is locked to the crossover peak of $|F_g=3\rangle$—$|F_e=2,3\rangle$ transitions of $^{85}$Rb $D_1$ line and used for nondestructive measurement during the evolution of the atoms between two microwave pulse interaction. The diameters for both of the laser beams are 6 mm while the powers are 12 mW and 50 μW for the resonant laser and the large detuning laser, respectively. The time sequences of the lasers are controlled by the acoustic optical modulators and are shown in Fig.2(b). The 6.834 GHz microwave signal is generated by a microwave synthesizer (RB-1, SpectraDynamics, Inc.) referred to a 5 MHz local oscillator (LO, BVA-8607, Oscilloquartz Corp.). With the use of a microwave switch (HMC-C019, Hittite Microwave Corp.), the microwave pulses with pulse width of 50 μs for the Ramsey interrogation are generated. A microwave phase shifter (HMC247, Hittite Microwave Corp.) is used to change the phase of the microwave pulse.

In the conventional Ramsey spectroscopy, there are two microwave pulses with phase of (0°, 0°) that interact with the atomic ensemble. The linewidth of the Ramsey fringe is decided by the evolution time of the atomic ensemble

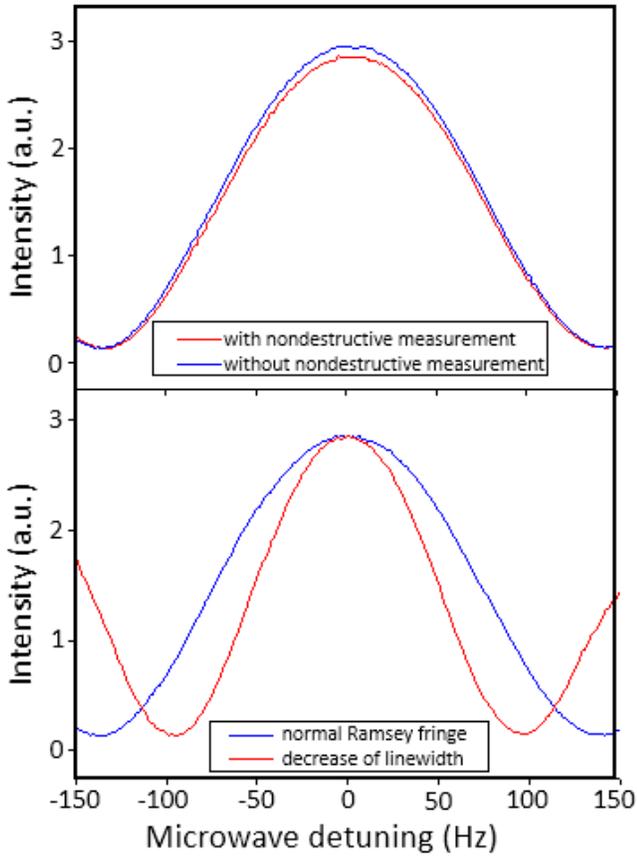

Fig.3 (color online) Demonstration of nondestructive measurement and decrease of Ramsey linewidth when *N*=2. (a) Central zone of Ramsey fringes with (red line) and without (blue line) nondestructive measurement. The signal intensity is reduced by 2% after nondestructive measurement is fulfilled. (b) Blue line is the central zone of normal Ramsey fringe (*N*=1) with linewidth of 142 Hz while red line is the fringe of *N*=2 with linewidth of 98 Hz.

between the two microwave pulses. We choose the evolution time as T=3.5 ms to obtain a high contrast Ramsey fringe. Fig.3.(a) (blue line) shows the Ramsey fringe with linewidth of 142 Hz. We then demonstrate the decrease of linewidth with *N*=2, i.e., we divide the Ramsey interrogation into two Ramsey sub-interrogation with evolution time of 1.7 ms. After evolving for 1.7 ms, a microwave pulse with phase of 90° is applied to interact with the atoms. Straight after that, a large detuning light probe the atomic phase and obtain $\varphi'_1$. Another microwave pulse with phase of -90° is then applied to cancel the interaction of the 90° microwave pulse. The atoms continue to evolve in the rest 1.7 ms. After the interactions of all the four microwave pulses (0°, 90°, -90°, 0°) and the laser pulse, we make the projective measurement and obtain the Ramsey fringe, as shown in Fig.3(a) (red line). Compared to the result obtained without nondestructive measurement (blue line), the signal intensity remains as large as 98%, which means the atomic coherence is well maintained after the extra introduction of the two microwave pulses (90°, -90°) and the nondestructive measurement.

By introducing the large detuning laser pulse during the free evolution, we peek the atomic phase $\varphi'_1$ in the middle of the interrogation. The phase $\varphi'_1$ is then fed back through the microwave phase shifter to change the fourth microwave pulses with phase of 90° -$\varphi'_1$. The phases of the four microwave pulses that interact with the atoms are (0°, 90°, -90°, 90° -$\varphi'_1$). The final Ramsey fringe is shown in Fig.3(b) (red line). The linewidth is about 98 Hz, which is about 2/3 of 142 Hz (blue line) obtained by 1/2*T*.

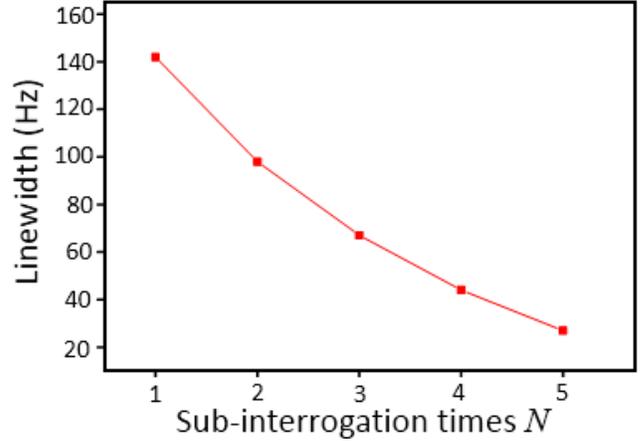

Fig.4 Demonstration of exponential decrease of Ramsey linewidth in the hot vapor cell atomic clock. The linewidths are 142, 98, 67, 44 and 27 Hz for N=1 to 5, respectively.

Since the coherent time of the hot vapor cell atomic clock is about 4 ms and it is much larger than the microwave pulse time of 0.05 ms, we continue to divide the Ramsey interrogation into N=3, 4, and 5 sub-interrogation. After every sub-interrogation, we probe the atomic phase nondestructively and change the phase of the microwave pulses in the following sub-interrogations. The phases for the microwave pulses are (0°, 90°, -90°, 90° -$\varphi'_1$, -90° +$\varphi'_1$, 90° -$\varphi'_2$), (0°, 90°, -90°, 90°-$\varphi'_1$, -90°+$\varphi'_1$, 90° -$\varphi'_2$, -90° +$\varphi'_2$, 90° -$\varphi'_3$ ), and (0°, 90°, -90°, 90°-$\varphi'_1$, -90°+$\varphi'_1$, 90° -$\varphi'_2$, -90° +$\varphi'_2$, 90° -$\varphi'_3$, -90° +$\varphi'_3$, 90° -$\varphi'_4$) for N=3, 4, and 5, respectively. As shown in Fig.4, after tracking the atomic phases in the evolution process, the linewidths for the final fringes are 67, 44, and 27 Hz for N=3, 4, and 5, respectively. The results fit well with the aforementioned equation $\Delta\nu = N/(2^N - 1) \cdot 1/2T$. Compared to the normal Ramsey interrogation with two microwave pulses, the linewidth of the Ramsey fringe here is effectively decreased by $N/(2^N-1)$, corresponding to the clock stability been improved by $N/(2^N-1)$. However, since we have fulfilled the measurement for *N* times, the signal to noise ratio for the final Ramsey fringe is then reduced by a factor of $1/\sqrt{N}$. With the narrow linewidth, the total result of the clock stability should be improved by about $N^{3/2}/(2^N-1)$ in short term stability.

In conclusion, we have proposed a new method to decrease the Ramsey linewidth of microwave atomic clock by peeping and tracking the coherent atomic phase through nondestructive measurement. The linewidth of Ramsey fringe decreases quickly as $N/(2^N-1)$ when the Ramsey interrogation is divided into *N* sub-interrogation. We demonstrate the proposal in a hot vapor cell atomic clock and obtain the narrow linewidth of Ramsey fringe down to 27 Hz after 5 sub-interrogations while its total interrogation time maintains 3.5 ms. We may use the method to improve the performance of atomic clock.

This work was supported by the National Nature Sciences Foundation of China under Grant No. 11504393 and No. 91536220; Foundation from Shanghai Science and Technology Committee under Grant No. 15YF1413400. Jinda Lin thanks Rong Wei for useful discussion.

*Corresponding author.

References
1. I. Rabi, J. Zacharias, S. Millman, and P. Kusch, Physical Review, 53, 318(1938).
2. N.F.Ramsey, Phys. Rev 78, 695(1950)
3. M.A. Lombardi; T.P. Heavner; S.R. Jefferts. "NIST Primary Frequency Standards and the Realization of the SI Second". Journal of Measurement Science. 2 (4): 74(2007)
4. Riehle F. 2004 Frequency Standards: Basics and Applications(Weinheim:Wiley-VCH) Chapter III,V.
5. A. Godone, S.Micalizio, and F. Levi, Phys. Rev. A 70, 023409(2004)
6. M.A.Kasevich, E.Riis, S.Chu, and R.G.DeVoe, Phys. Rev. Lett. 63, 612(1989)
7. Yoav Sagi, Ido Almog, and Nir Davidson, Phys. Rev. Lett. 105, 053201 (2010)
8. C.Deutsch, F.Ramirez-Martinez, C.Lacroute, F.Reinhard, T.Schneider, J.N.Fuchs, F.Piechon, F.Laloe, J.Reichel, and P.Rosenbusch, Phys. Rev. Lett. 105,020401(2010)
9. G. Kleine Buning,_ J. Will, W. Ertmer, E. Rasel, J. Arlt, and C. Klempt, F. Ramirez-Martinez, F. Piechon, and P. Rosenbusch, Phys. Rev. Lett. 106. 240801(2011)
10. R.Kohlhaas, A. Bertoldi, E.Cantin, A.Aspect, A.Landragin, and P.Bouyer, Phys. Rev. X 5,021011
11. N. Shiga and M Takeuchi, New J. Phys. 14, 023034 (2012)
12. N.Shiga, M. Mizuno, K.Kido, P. Phoonthong and K. Okada, New J. Phys. 16, 073029(2014)
13. G.W.Biedermann, K. Takase, X.Wu, L.Deslauriers, S.Roy, and M.A. Kasevich, Phys. Rev. Lett. 111,170802 (2013)
14. F. Bloch, Phys. Rev. 70 460 (1946)
15. S. Micalizio, C. E. Calosso, A. Godone, and F. Levi, Metrologia. **49,** 425 (2012).
16. Jinda Lin, Jianliao Deng, Yisheng Ma, Huijuan He, and Yuzhu Wang, Opt. Lett. 37, 5036 (2012)